\documentclass[pre,twoside,twocolumn,superscriptaddress,floatfix,showpacs]{revtex4-1}

\usepackage{graphicx,amssymb,amsmath,epsfig,color,enumerate,bm,psfrag}
\usepackage{afterpage,placeins,float}
\usepackage{subcaption}

\begin{document}
 \title{Ergodicity recovery of random walk\\ in heterogeneous disordered media}
  \author{Liang Luo}
   \email{luoliang@mail.hzau.edu.cn}
 \affiliation{Department of Physics, Huazhong Agricultural University, Wuhan 430070, China}
 \affiliation{Institute of Applied Physics, Huazhong Agricultural University, Wuhan 430070, China}
 \author{Ming Yi}
 \affiliation{School of Mathematics and Physics, China University of Geosciences, Wuhan 430074, China}

\begin{abstract}
Significant and persistent trajectory-to-trajectory variance are commonly observed in the particle tracking experiments, which have become a major challenge for the experiment data analysis. In this theoretical paper, we investigate the ergodicity recovery behavior, which helps to clarify the origin and the convergence of trajectory-to-trajectory fluctuation in various heterogeneous disordered media. The concepts of self-averaging and ergodicity are revisited in the context of trajectory analysis. The slow ergodicity recovery and the non-Gaussian diffusion in the annealed disordered media are shown as the consequences of the central limit theorem in different situations. The strange ergodicity recovery behavior is reported in the quenched disordered case, which arises from a localization mechanism. The first-passage approach is introduced to the ergodicity analysis for this case, of which the central limit theorem can be employed and the ergodicity is recovered in the length scale of diffusivity correlation.
\end{abstract}


\maketitle

\section{Introduction}

Particle tracking experiments on various disordered systems, including the living cells\cite{li15,he16,munder16,li18}, colloidal\cite{sentjabrskaja16,ning19} and granular\cite{kou17} systems have provided numerous trajectories with rich dynamic details. It has been utilized to infer the latent dynamics of the tracer\cite{manzo15} and also the disordered feature of the environments\cite{munder16,luo18}, which calls for careful statistics analysis\cite{barkai12,metzler14} on the random walks.

The commonly observed significant trajectory-to-trajectory variance is one of the major challenges in the trajectory analysis. Due to the stochastic nature of random walk, it exists even in the normal Brownian motion in the homogeneous media. In the simple case the fluctuation is depressed in longer trajectories. The ergodicity hence recovers. It is observed in experiments, however, the trajectory-to-trajectory variance can sustain in disordered media over the whole experiments\cite{li15,he16}, which has been considered as a consequence of the heterogeneity of dynamics in the media. In the case of strong disorder, the heterogeneity leads to sub-diffusive continuous time random walk(CTRW)\cite{montroll75,luo14}. The ergodicity would not recover in such case\cite{bel05,he08}. In the case of the moderate heterogeneity, one may also observe slow ergodicity recovery while the random diffusivity correlates along the trajectory, in which case the non-Gaussian diffusion has been intensively studied\cite{wang09,he16,jeon16,liu17}. In the recent study on the non-Gaussian diffusion\cite{luo18,luo19}, a localization mechanism is discovered in the quenched disordered media with locally correlated diffusivity. The population splitting\cite{cherstvy13,luo19} due to the localization introduces strange and ultra-slow recovery of ergodicity. The similar behavior has also been reported in the molecular dynamics simulation\cite{jeon16}. It is currently unclear whether and how the ergodicity recovers in the quenched disordered case\cite{luo19,guo18}.

In this paper, we crystalize the idea of self-averaging and ergodicity recovery by the model study in the fashion of experiment trajectory analysis, where the trap model\cite{machta81,haus87, luo14} is employed as a theoretical framework containing the random walk in homogeneous media, in the annealed disordered media with temporally correlated diffusivity and in the quenched disordered media with spatially correlated diffusivity. One will see that the central limit theorem (CLT) plays a key role in most of the material dealt in this paper, which is also connected to the non-Gaussian diffusion. We suggest that the first-passage time would be a proper observable for the  investigation on the ergodicity recovery in the quenched disordered system.

This paper is arranged as follows. We introduce in Sec. \ref{sec:sa} the concept of ergodicity in trajectories. The simple homogeneous case is revisited as an example to show in which sense we say the trajectories are similar with each other. In Sec. \ref{sec:annealed}, we turn to the annealed disordered system. One can see how the self-averaging leads to the ergodicity recovery when the observation time is much longer than the relaxation time of the diffusivity. In Sec. \ref{sec:quenched}, we study the more complicated case of quenched disorder, where the self-averaging are realized by sampling large region of the static disordered landscape. The first-passage approach is employed here. Section \ref{sec:disc} and Section \ref{sec:sum} are the discussion and the summary. The study on the non-Gaussian distribution of the displacement is shown in Appendix \ref{app1}. The simulation details are provided in Appendix \ref{app2}.

\section{Ergodicity of random walk in homogeneous medium: revisited}
\label{sec:sa}
Let's consider the trap dynamics on a two-dimensional square lattice with lattice constant $a$, of which a particle jumps from site $i$ evenly to its nearest-neighbour site $j$ with the transition rate $w$. The stochastic processes can be then considered as a normal random walk on lattice subordinated by a time series defined by the waiting time for each jump, $\{t_i\}$, where $t_i$ follows exponential distribution by
\begin{equation}
P(t_i=t)=4w\exp\left(-4w t\right).
\end{equation}
The transition rate $w$ can be defined in different ways to model the walks in various environments.
In the case of diffusion in homogeneous media, one can assign constant $w$ for all the jumps. The random walk turns to be the normal Brownian motion in the long time limit. In more complicated cases with heterogeneity in dynamics, $w$ is a random variable fluctuating over time or depending on site, which introduces anomalies such as non-Gaussian displacement distribution and sub-diffusion. We consider in this section the homogeneous case with constant $w$.

\begin{figure}
\includegraphics[width=.9\linewidth]{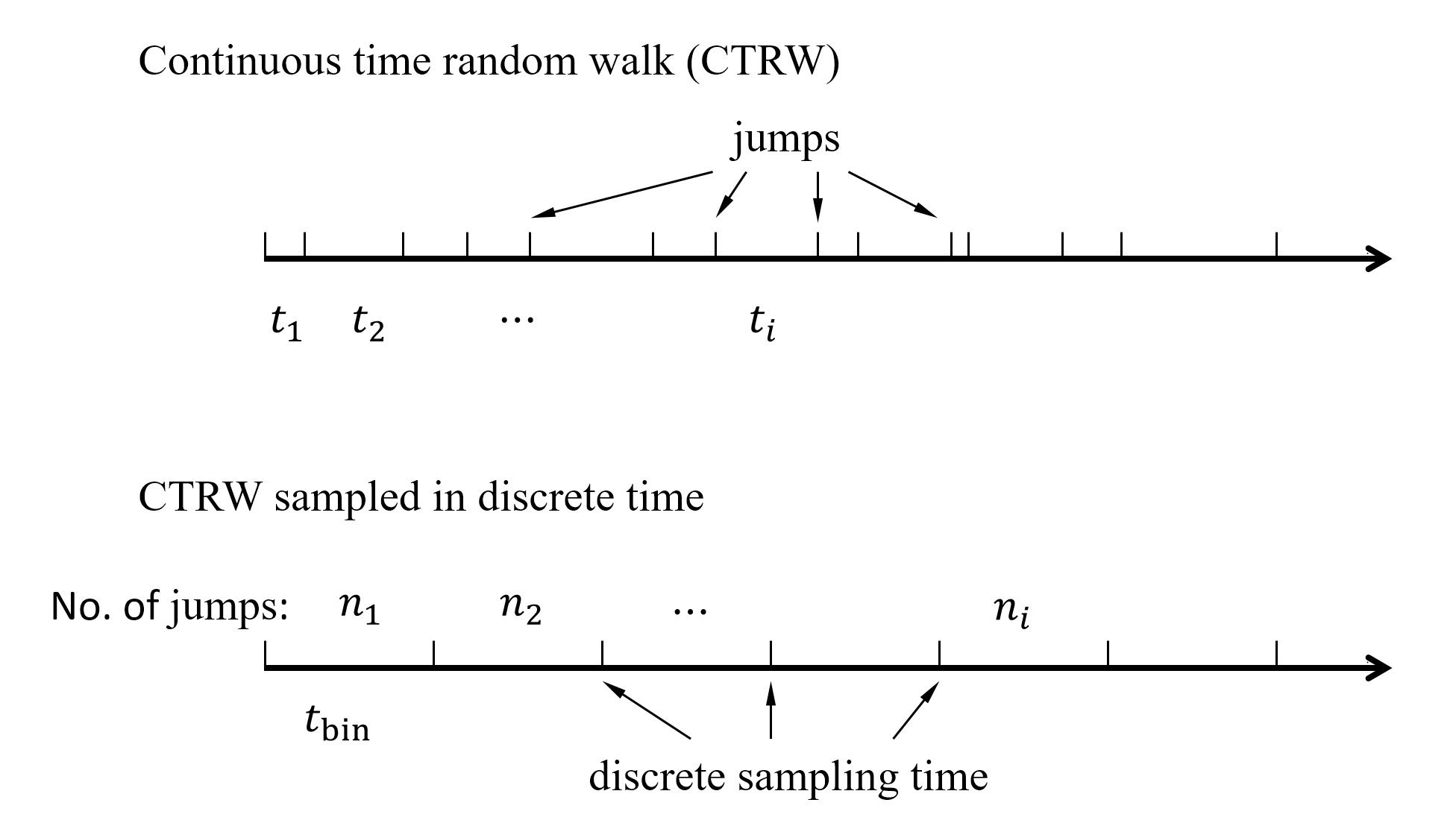}
\caption{\label{fig:ctrw} The time course of continuous time random walk and that sampled in discrete time.}
\end{figure}

The dynamics defined above is often called continuous time random walk (CTRW) in literature, which is discrete in space and continuous in time. The particle tracking experiment data is, however, in a different style that the particle positions are recorded by fixed time interval. For better guidance to the experiment data analysis, the trajectories from the trap dynamics are discretized into the series of the particle positions $\{{\bf x}_i\}$ with the constant time interval $t_{\text{bin}}=t_i-t_{i-1}$, as shown in Fig. \ref{fig:ctrw}. One can introduce the displacement increment
\begin{equation}
\label{eq:xi}
\xi_i={\bf x}_i-{\bf x}_{i-1}
\end{equation}
in each time interval. Noting the successive jumps of CTRW have no direction correlation, i.e. $\left<\xi_i\cdot\xi_j\right>=\delta_{ij}$ with $\delta_{ij}$ the Kronecker delta, one can see
\begin{equation}
\xi_i^2\simeq n_i a^2, \text{ for } n_i\gg1,
\end{equation}
where $n_i$ is the number of jumps in time interval $(t_{i-1},t_i)$.
In the homogeneous case, the jumps happen in the time scale $t_0\sim(4w)^{-1}$. In the case that $t_{\text{bin}}\gg t_0$,  multiple jumps happen in one time interval and $\vert\xi\vert\gg a$. The lattice feature hence leaves the discretized trajectory. It becomes a random walk continuous in space and discrete in time.

 In the simulation on the homogeneous case, we are sure the trajectories are {\it similar}, since the same and constant transition rate. It is, however, not trivial to verify the similarity of the trajectories obtained in experiments, where the underlying mechanism is usually unknown. For more rigorous analysis, one may turn to the concept of {\it ergodicity}, which refers that the observable averaged along each trajectory equals to that averaged over the ensemble of the trajectories.
The most commonly used observable in the particle tracking experiments is the time-averaged mean squared displacement (TAMSD)\cite{he08,metzler14}, which is a trajectory-wise version of mean squared displacement (MSD). The TAMSD averages the square of the head-to-tail displacement of the short segments from single trajectory by
\begin{equation}
\overline{\delta^2}(\Delta_t,T)=\frac{1}{T-\Delta_t}\int_0^{T-\Delta_t} dt\;\delta^2(t,\Delta_t),
\end{equation}
where $\Delta_t$ is the time duration of the short segments, $T$ is the duration of the long trajectory, and $\delta^2(t,\Delta_t)=\vert{\bf x}(t+\Delta_t)-{\bf x}(t)\vert^2$. In the time-discretized version,
\begin{equation}
\label{eq:tamsd}
\overline{\delta^2}(\Delta_t,T)=\frac{1}{M}\sum_{k=1}^{M}\delta_k^2(\Delta_t).
\end{equation}
where
\begin{equation}
\label{eq:dx}
\delta_k(\Delta_t)={\bf x}(t_k+\Delta_t)-{\bf x}(t_k)
\end{equation}
is calculated for the segment initiated at time $t_k$, which is determined by the time lag $t_{\text{lag}}=t_k-t_{k-1}$. To avoid the correlation due to the overlap of segments, it is required that $t_{\text{lag}}\ge\Delta_t$ . The equality is adopted here to utilise all the frames recorded in the trajectories. There are in total $M=(T-\Delta_t)/\Delta_t$ segments sampled from the trajectory.

We estimate here the distribution of trajectory-wise TAMSD defined by Eq.(\ref{eq:tamsd}).
One may note
\begin{equation}
\label{eq:disp1}
\delta_k(\Delta_t)=\sum_{i=1}^{N_q}\xi_{k+i},
\end{equation}
with $N_q=\Delta_t/t_{\text{bin}}$.
The squared displacement hence follows
\begin{equation}
\delta_k^2(\Delta_t)=\sum_{i,i'}\xi_{k+i}\cdot\xi_{k+i'}=\sum_{i=1}^{N_q}\xi_{k+i}^2,
\end{equation}
where$\left<\xi_i\cdot\xi_j\right>=\delta_{ij}$ is applied for the second equality. The trajectory-wise TAMSD is then given by
\begin{equation}
\label{eq:tamsd1}
\overline{\delta^2}(\Delta_t,T)=\frac{1}{M}\sum_{k=1}^{M}\sum_{i=1}^{N_q}\xi_{k+i}^2=\frac{N_q}{N}\sum_{j=1}^{N}\xi_j^2=N_q \overline{\xi^2},
\end{equation}
where $\overline{\xi^2}$ is averaged over the trajectory and $N=Mq=(T-\Delta_t)/t_{\text{bin}}$ is roughly the number of the frames of the whole trajectory when $\Delta_t\ll T$.
The problem turns to estimation the distribution of $\overline{\xi^2}$.
Since the independent jumps on the lattice follows the constant rate $w$, the number of the jumps $n_i$ in the time interval $(t_{i-1},t_i)$ follows Poisson distribution with the expectation $n_{b}=4wt_{\text{bin}}$, i.e.
\begin{equation}
\label{eq:poisson}
P(n_i=z)=\frac{{n_{b}}^z}{z!}e^{-n_{b}}.
\end{equation}
One may note the displacement distribution of two-dimensional isotropic random walk of $n_i$ steps follows the Gaussian distribution by
\begin{equation}
P(x_i,y_i\vert n_i)\simeq \frac{1}{\pi n_i a^2}\exp\left(-\frac{x_i^2+y_i^2}{n_i a^2}\right), \text{ for }n_i\gg1.
\end{equation}
Noting $\xi_i^2\equiv x_i^2+y_i^2$, we can see it follows the exponential distribution by
\begin{equation}
\label{eq:exp}
P(\xi_i^2=z\vert n_i)=\frac{1}{n_i a^2}\exp\left({-z/n_ia^2}\right).
\end{equation}
From Eq. (\ref{eq:poisson}) and Eq. (\ref{eq:exp}), we arrive at the expectation $\left<\xi^2\right>=n_b a^2$,
and the variance $\left<\left[\xi^2\right]^2\right>=\left<\xi^2\right>^2$.
Since Eq. (\ref{eq:tamsd1}) shows that $\overline{\delta^2}/N_q$ is the mean value of $\xi^2$ of the population of $N=(T-\Delta_t)/t_{\text{bin}}$. The CLT thus suggests
\begin{equation}
\label{eq:clthomo}
\overline{\delta^2}(\Delta_t,T)/N_q\xrightarrow{d}\mathcal{N}\left(\left<\xi^2\right>,\left<\left[\xi^2\right]^2\right>/N\right),
\end{equation}
where $\mathcal{N}(\mu,\sigma^2)$ denotes the Gaussian distribution of the expectation $\mu$ and the variance $\sigma^2$.
In this case, $\mu=\left<\xi^2\right>$ and $\sigma^2=\left<\left[\xi^2\right]^2\right>/N$, i.e.
It hence gives
\begin{equation}
\label{eq:homo1}
\left<\overline{\delta^2}(\Delta_t,T)\right>=N_q \left<\xi^2\right>=4wa^2\Delta_t
\end{equation}
and
\begin{equation}
\label{eq:homo2}
\left<\left[\overline{\delta^2}(\Delta_t,T)\right]^2\right>=N_q^2\left<\left[\xi^2\right]^2\right>/N=\left(4wa^2\Delta_t\right)^2 \left[(T-\Delta_t)/t_{\text{bin}}\right]^{-1},
\end{equation}
where $\left<\cdot\right>$ denotes the average over trajectories.
In the trajectory analysis practice, $\Delta_t$ should be always kept much smaller than $T$. Or one may encounter abnormal large fluctuation in TAMSD when $(T-\Delta_t)\sim t_{\text{bin}}$. $(T-\Delta_t)/t_{\text{bin}}$ is hence roughly the number of the total frames of the trajectory. One can see the variance of $\overline{\delta^2}$ is suppressed by the self-averaging among the frames in each trajectory. The mean diffusivity along each trajectory, $\overline{D}\equiv\overline{\delta^2}(\Delta_t,T)/4\Delta_t$, then converges to its ensemble expectation $\left<D\right>=wa^2$. This is the simplest example showing how the ergodicity recovers in long random walks.

The so-called ergodicity breaking parameter\cite{he08} is introduced as the square of relative standard deviation of $\overline{\delta^2}$ by
\begin{equation}
\label{eq:eb}
\text{EB}=\frac{\left<\left[\overline{\delta^2}\right]^2-\left<\overline{\delta^2}\right>^2\right>}{\left<\overline{\delta^2}\right>^2}.
\end{equation}
It has been widely employed in the trajectory analysis\cite{cherstvy13,metzler14,jeon14,jeon16}
One can easily read from Eq.(\ref{eq:homo1}) and Eq.(\ref{eq:homo2}) that $\text{EB}\simeq t_{\text{bin}}/T=1/N$ in the homogeneous case.

\section{Ergodicity recovery of random walk in the annealed disordered media}
\label{sec:annealed}

In this section, we study the case that the instantaneous diffusivity fluctuates along the trajectory, which is commonly observed in experiments. The classical CTRW\cite{montroll75,haus87,bouchaud90} offers a way to capture the feature by sampling the waiting time for every jump from a non-exponential distribution $P(t)$, which has been a successful model to explain the anomalies in sub-diffusion. In the framework of trap dynamics defined above, it is equivalent to the case that the transition rate $w$ is resampled after each jump from the distribution $P(w)$\cite{luo15}.

It is realized in recent years that the diffusivity can be a stochastic process {\it independent} of the jumps in the case with certain latent dynamics, such as the fluctuating configuration of the protein tracer or the transient interaction between the protein and the cell membrane\cite{manzo15}. To include this case, one may modify the classical CTRW model by introducing an additional latent dynamics, of which $w$ is resampled from the distribution $P(w)$ by a rate $w_{D}=1/ t_{\text{D}}$. $w$ is then correlated in the time scale $t_{\text{D}}$. When the correlation time of the diffusivity, $ t_{\text{D}}$, is in a moderate scale, i.e. $\Delta_t< t_{\text{D}}<T$, one may observe the non-Gaussian distribution of the head-to-tail displacement $\vert\delta\vert=\vert {\bf x}(t+\Delta_t)-{\bf x}(t)\vert$ of short segments. Noting that the latent dynamics fluctuates over time and is independent of the particle location, one can see it is an {\it annealed} model for the non-Gaussian diffusion, which can be also understood as a lattice version of the diffusing diffusivity model introduced by Chubynsky and Slater\cite{slater14}. In this section, we study the case that $w$ follows the generalized Gamma distribution by
\begin{equation}
\label{eq:pw}
P(w)=\alpha w^{\alpha-1}\exp\left(-w^{\alpha}\right),
\end{equation}
where the parameter $\alpha$ modulate the heterogeneous level of the dynamics\cite{luo18,sposini18,luo19}. In the case $\alpha\rightarrow\infty$, $P(w)$ converges to a sharp peak at $w=1$, which turns back to the homogeneous case studied in Sec. \ref{sec:sa}. The displacement distribution is in general non-Gaussian for $\alpha<\infty$. One can find the analysis on the non-Gaussian behavior in Appendix \ref{app1}.

\begin{figure}
  \centering
  \includegraphics[width=.9\linewidth]{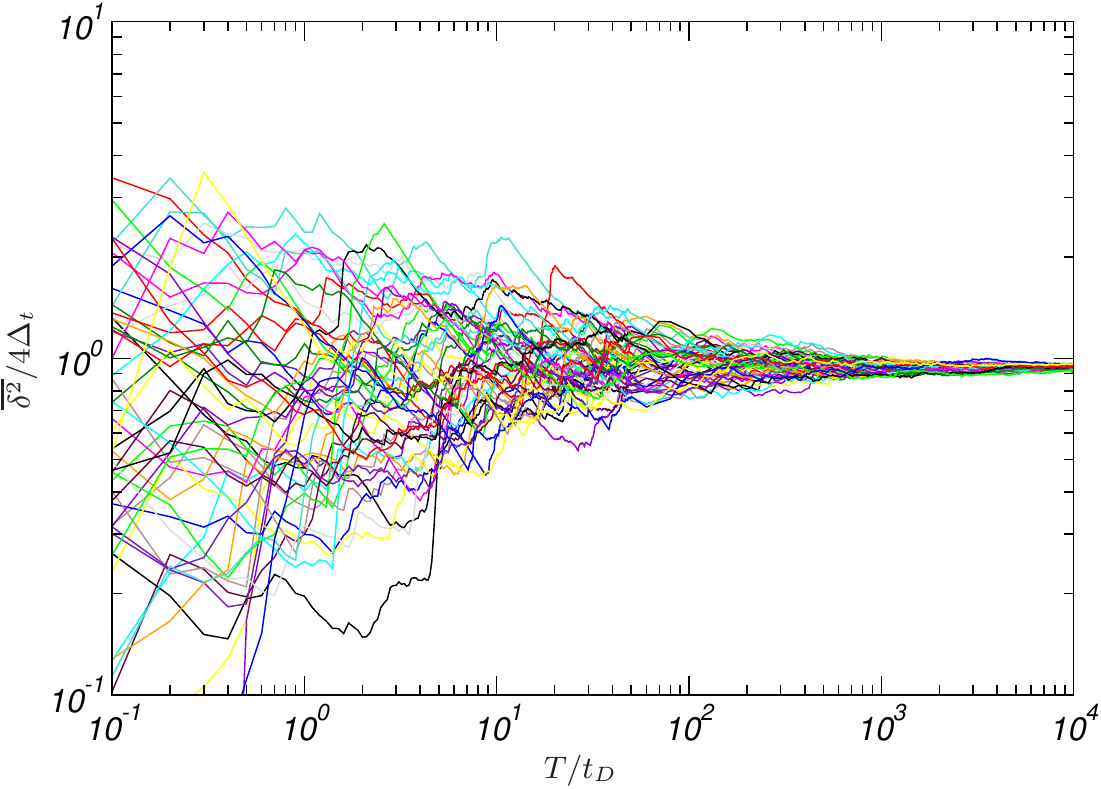}
  \caption{ The rescaled trajectory-wise TAMSD $\overline{\delta^2}/4\Delta_t$ versus the rescaled observation time $T/t_D$ in the annealed disordered case. It contains $50$ typical trajectories with $\alpha=1.2$, $t_D=200$ and $\Delta_t=16$. }
  \label{fig:tamsd_annealed}
\end{figure}

In this annealed disordered model, the trajectory-to-trajectory variance sustains in the time scale of $\tau_D$, which vanishes for longer observation time. Figure \ref{fig:tamsd_annealed} presents $50$ trajectory-wise TAMSD $\overline{\delta^2}/4\Delta_t$ for increasing observation time $T$. We investigate the ergodicity recovery behavior below.

One can start from Eq.(\ref{eq:tamsd1})
\begin{equation}
\overline{\delta^2}(\Delta_t,T)=\frac{N_q}{N}\sum_{j=1}^{N}\xi_j^2.
\end{equation}
where $N=T/t_{\text{bin}}$ and $N_q=\Delta_t/t_{\text{bin}}$.
In the short time limit with $T< t_{\text{D}}$, the diffusivity $w$ is roughly unchanged in each concerned trajectory. The argument for Eq. (\ref{eq:clthomo}) may also apply here, which suggests
\begin{equation}
\overline{\delta^2}(\Delta_t,T\vert w)\xrightarrow{d}\mathcal{N}\left(\mu,\sigma^2\right),
\end{equation}
with the expectation $\mu=N_q\left<\xi^{2}\right>=4wa^2\Delta_t$ and the variance $\sigma^2=N_q^2\left<\left[\xi^{2}\right]^2\right>/N=(4wa^2\Delta_t)^2t_{\text{bin}}/T$.
It is quite similar to Eq.(\ref{eq:homo1}) and Eq.(\ref{eq:homo2}) in the homogeneous case, except that $w$ is a random variable.
Considering the $t_{\text{bin}}\ll T$ limit, the distribution becomes a sharp peak around $\overline{\delta^2}=4wa^2\Delta_t$.
In the trajectory ensemble, the marginal distribution gives
\begin{eqnarray}
P(\overline{\delta^2}=x)&=&\int_0^\infty dw\; P(\overline{\delta^2}=x\vert w)P(w)\nonumber\\
&\simeq&\int_0^\infty dw\; P(w)\delta(x-4wa^2\Delta_t).
\end{eqnarray}
The Dirac-$\delta$ function appears in the second line.
In this case, the trajectory-to-trajectory variance of $\overline{\delta^2}$ is mainly contributed by the random diffusivity $w$, which is highly related to the non-Gaussian diffusion discussed in the Appendix \ref{app1}.

\begin{figure}
  \centering
  \includegraphics[width=.9\linewidth]{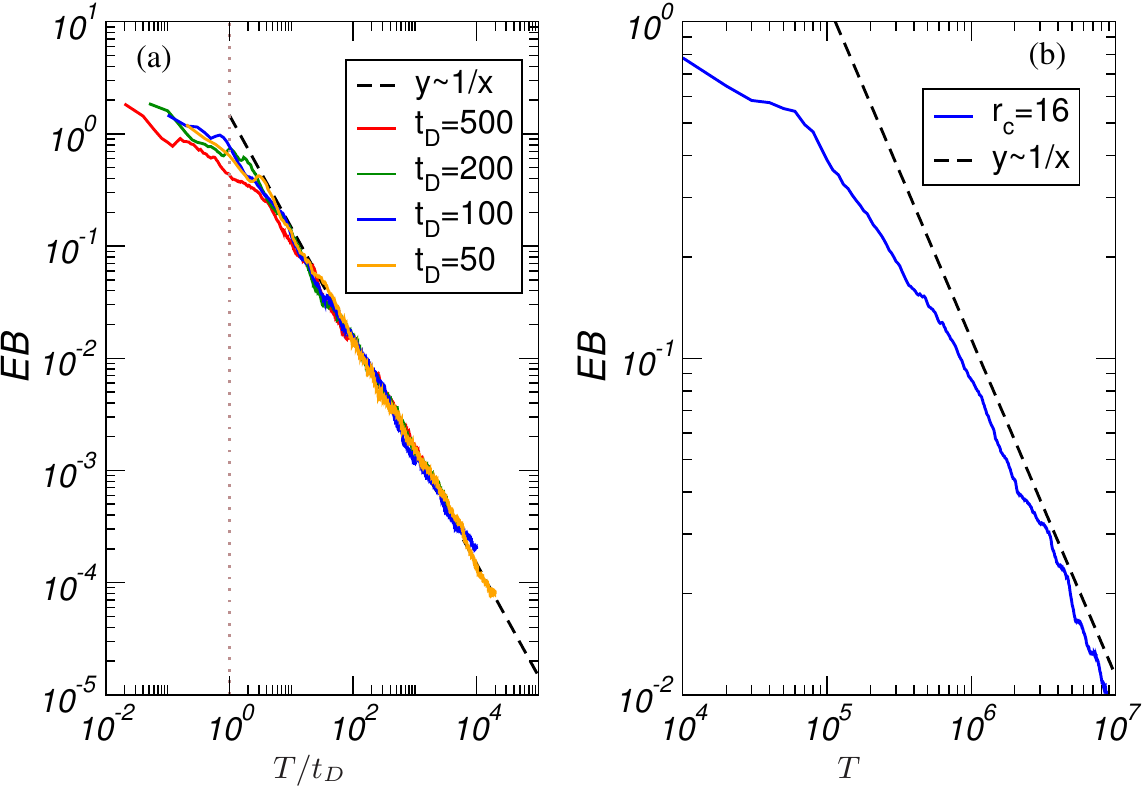}
  \caption{The ergodicity breaking parameter in the disordered cases with $\alpha=1.2$. (a) The annealed cases with various $ t_{\text{D}}$ (colored solid lines).  The dash line indicates the ergodicity recovery by $EB\sim (T/ t_{\text{D}})^{-1}$. (b) The quenched case with $r_c=16$.
}
  \label{fig:eb}
\end{figure}

In the long time limit with $T\gg t_{\text{D}}$, one can regroup the summation terms in Eq.(\ref{eq:tamsd1}) by the time interval $ t_{\text{D}}$ by
\begin{equation}
\overline{\delta^2}(\Delta_t,T)=\frac{N_q}{N}\sum_{j=1}^{N_t}\delta_j^2( t_{\text{D}}),
\end{equation}
where
\begin{equation}
\delta_j^2( t_{\text{D}})=\sum_{i=1}^{N_p}\xi^2_{(j-1)N_p+i}
\end{equation}
with $N_p= t_{\text{D}}/t_{\text{bin}}$ and $N_t=T/ t_{\text{D}}$.
Assuming the $\xi^2$ in one subgroup follows the same $w$, one can use again the CLT to get
\begin{equation}
\delta^{2}_j( t_{\text{D}})\xrightarrow{d}\mathcal{N}\left(\mu_j,\sigma_j^2\right),
\end{equation}
where $\mu_j=4w_j a^2 t_{\text{D}}$ and $\sigma^2_j=(4w_ja^2)^2 t_{\text{bin}} t_{\text{D}}$.
$\overline{\delta^2}(\Delta_t,T)$ is then a summation over Gaussian distributed variables. One can immediately get
\begin{equation}
\overline{\delta^2}(\Delta_t,T)\xrightarrow{d}\mathcal{N}\left(\mu,\sigma^2\right)
\end{equation}
with $\mu=4a^2\Delta_t \overline{w}$ and $\sigma^2=(4a^2\Delta_t)^2 \overline{w^2}t_{\text{bin}}/T$.
We would like to call the reader's attention here $\overline{w}=\frac{1}{N_t}\sum_{j=1}^{N_t}w_j$ and  $\overline{w^2}=\frac{1}{N_t}\sum_{j=1}^{N_t}w_j^2$ are both random variables depending on the realization of $\{w_j\}$.
In the case $N=T/t_{\text{bin}}\gg 1$,  $\sigma^2$ vanishes for $T\gg t_{\text{bin}}$. One can assume $\overline{\delta^2}=\mu=4a^2\Delta_t \overline{w}$.
For $N_t\gg1$, we take the aid from the CLT again to get
\begin{equation}
\overline{w}\xrightarrow{d}\mathcal{N}\left(\left<w\right>,\left<w^2\right>/N_t\right),
\end{equation}
where $\left<w\right>=\int_0^{\infty}dw\; wP(w)$ and $\left<w^2\right>=\int_0^{\infty}dw\;w^2P(w)$.
The distribution of $\overline{\delta^2}$ is eventually obtained as
\begin{equation}
\overline{\delta^2}\xrightarrow{d}\mathcal{N}\left(4a^2\Delta_t\left<w\right>,(4a^2\Delta_t)^2\left<w^2\right>/N_t\right).
\end{equation}
The normalized variance of $\overline{\delta^2}$ (EB parameter) then vanishes for $T\gg t_{\text{D}}$ by
\begin{equation}
EB=\frac{ t_{\text{D}}}{T}\frac{\left<w^2\right>}{\left<w\right>^2},
\end{equation}
which is confirmed by the simulation data as shown in Fig. \ref{fig:eb}(a).

\section{Ergodicity recovery of random walk in the quenched disordered media}
\label{sec:quenched}

In this section, we study the random walk in the quenched disordered media, of which the diffusivity depends on the local structures of the environments. In the case that the structures relax in quite long time scale, the local diffusivity can be assumed unchanged over the experiments.  In the framework of trap dynamics, the quenched trap model (QTM)\cite{machta81} assigns the random transition rates $\{w_i\}$ to sites $\{i\}$ in the lattice.
In the experiment with spatial resolution high enough to reveal the local structures, the measured local diffusivity is usually correlated in the scale of the structure size. To include the locally correlated dynamics, we study the trap dynamics on the extreme landscape\cite{luo18,luo19}, which is an extension of QTM with the locally correlated $\{w_i\}$.

The extreme landscape $\{v_i\}$ is generated by the extreme statistics as follows. First to generate the uncorrelated auxiliary potential $\{u_i\}$ following the distribution with finite expectation and variance, such as the exponential distribution $P(u_i=u)=u_0^{-1}\exp(u/u_0)$ with $u<0$. The local minimal value of $\{u_i\}$ is then assigned to $v_i$, i.e. $v_i=\min\{u_j\vert r_{ij}<r_c\}$.  Each minimal value controls an area of the landscape, called ``extreme basin''. $\{v_i\}$ is identity in the extreme basin, of which the radius is constrained by $r_c$. Since $v_i$ is the minimal value of a set of independent $u_j$, in the case $r_c^2\gg1$ it follows the Gumbel distribution by
\begin{equation}
P(v_i=v)=\exp(v-v_0-\exp(v-v_0)).
\end{equation}
The trap dynamics gives the transition rate as $w_i=w_0\exp(v_i/\alpha)$.
Setting $w_0=\exp(-v_0/\alpha)$, one can see $w_i$ follows the generalized Gamma distribution given by Eq. (\ref{eq:pw}). In the low temperature case with $\alpha<1$, the distribution of the typical waiting time $\tau_i=(4w_i)^{-1}$ is with heavy tail. Sub-diffusion is then the consequence. It is well known that the ergodicity is absent in this case\cite{bel05,he08}. In the $\alpha>1$ case, the population splitting is introduced by a localization mechanism. The trajectory-to-trajectory fluctuation sustains till all the particles exit the localized state, which leads to quite slow ergodicity recovery.

\begin{figure}
  \centering
  \includegraphics[width=.9\linewidth]{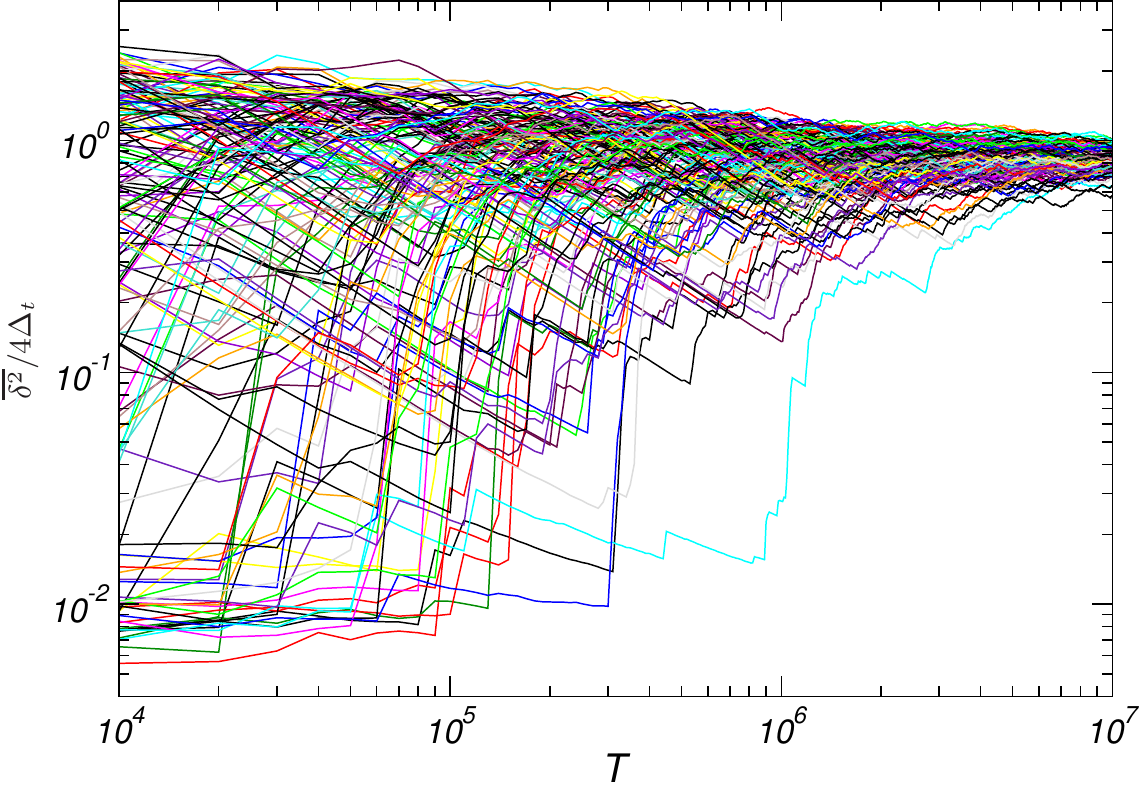}
  \caption{ The rescaled trajectory-wise TAMSD $\overline{\delta^2}/4\Delta_t$ versus the observation time $T$ in the quenched disordered case, where $\Delta_t=16$. It contains $160$ typical trajectories on the sample with $\alpha=1.2$, $r_c=16$. }
  \label{fig:tamsd_quenched}
\end{figure}

Without lose of generality, we investigate the ergodicity recovery in the quenched disordered case with $\alpha=1.2$. The extensive simulation is performed for 160 trajectories of quite long time ($T=10^7$), to guarantee the landscape is fully sampled. Figure \ref{fig:eb}(b) shows the EB parameter. It decreases quite slow for longer observation time $T$, and does not follow the $EB\sim 1/T$ rule even for very long $T$. It is more clearly illustrated by the rescaled trajectory-wise TAMSD $\overline{\delta^2}/4\Delta_t$, as shown in Fig. \ref{fig:tamsd_quenched}.  Compared with the annealed case, the most significant feature is that the TAMSDs of some trajectories are pinned at very small value for long time, before large hops bringing them to the expected value. These TAMSDs are contributed by the trajectories initially trapped in the slowest area of the landscape. They eventually enters the mobile area, which is remarked by the large hops. The waiting time for the escaping from the slowest area couples the local diffusivity, which can span several magnitudes as shown in the figure. Noting that the diffusivity is roughly constant (and hence strongly correlated) when the particle is localized in the deepest traps, one can see the CLT approach for independent variables is applicable only for quite large $T$. (For the $160$ simulated trajectories, $T>10^6$. )

\begin{figure}
  \centering
  \includegraphics[width=.9\linewidth]{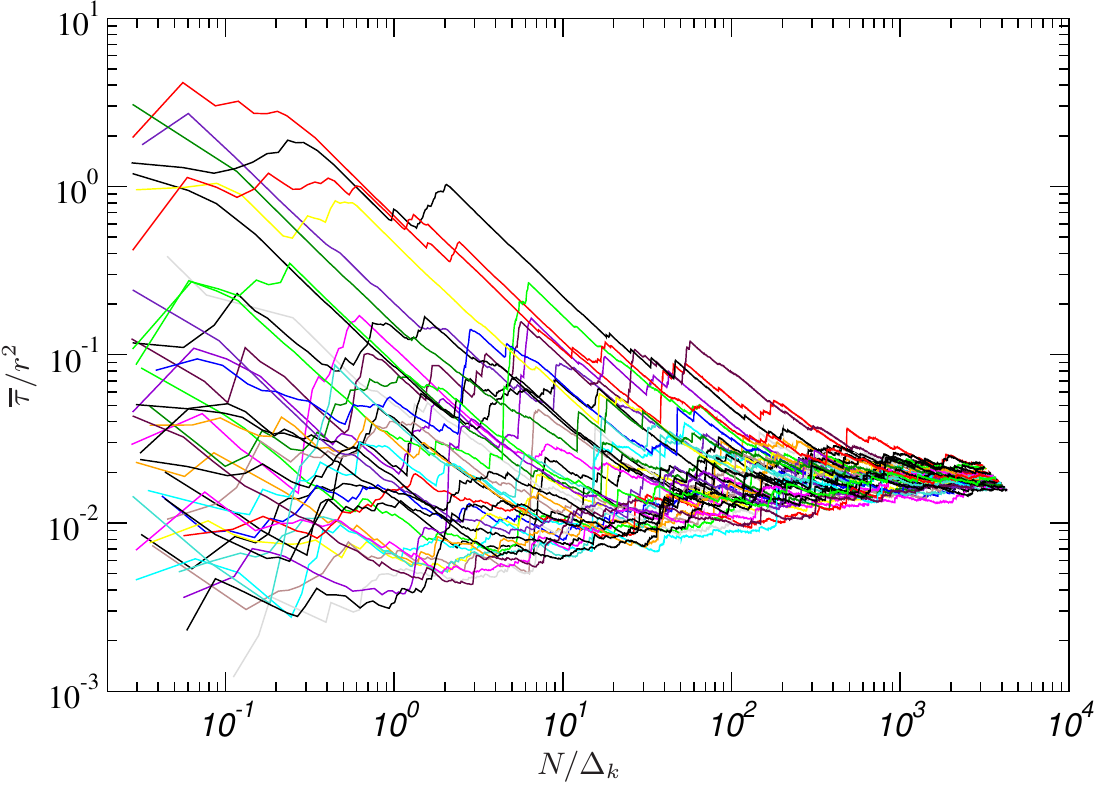}
  \caption{The mean first-passage time of $40$ trajectories on the sample with $\alpha=1.2$,$r_c=16$.
  }
  \label{fig:fpt_quenched}
\end{figure}

To clarify the self-averaging behavior in the quenched case, one may turn to another observable - the trajectory-wise mean first-passage time (FPT). In the first-passage approach, the trajectory $\{{\bf x}(t)\}$ is divided into segments determined by the successive first-passage events to radius $r$ at time $\{t_k\}$. The events can be formally defined by the conditions
\begin{equation}
\vert {\bf x}(t)-{\bf x}(t_k)\vert<r, \; t_k<t<t_{k+1},\nonumber
\end{equation}
and
\begin{equation}
\vert {\bf x}(t_{k+1})-{\bf x}(t_k)\vert\ge r,\nonumber
\end{equation}
The first-passage time $\tau$ is then defined as $\tau_k=t_{k+1}-t_k$, which apparently depends on $r$.
By the definitions, one can see the duality between the square displacement and the first-passage time. The former one concerns the fluctuating displacements of the segments with the fixed time duration. The later one concerns the fluctuating time durations of the segments with the fixed head-to-tail distance. The trajectory-wise mean first-passage time can be defined along a trajectory of $N$ successive first-passage segments by
\begin{equation}
\label{eq:mfpt}
\overline{\tau}=\frac{1}{N}\sum_{k=1}^N \tau_k.
\end{equation}
Figure \ref{fig:fpt_quenched} shows the mean first-passage time along $40$ simulated trajectories, which converge to the expectation for large $N$.

\begin{figure}
  \centering
  \includegraphics[width=.9\linewidth]{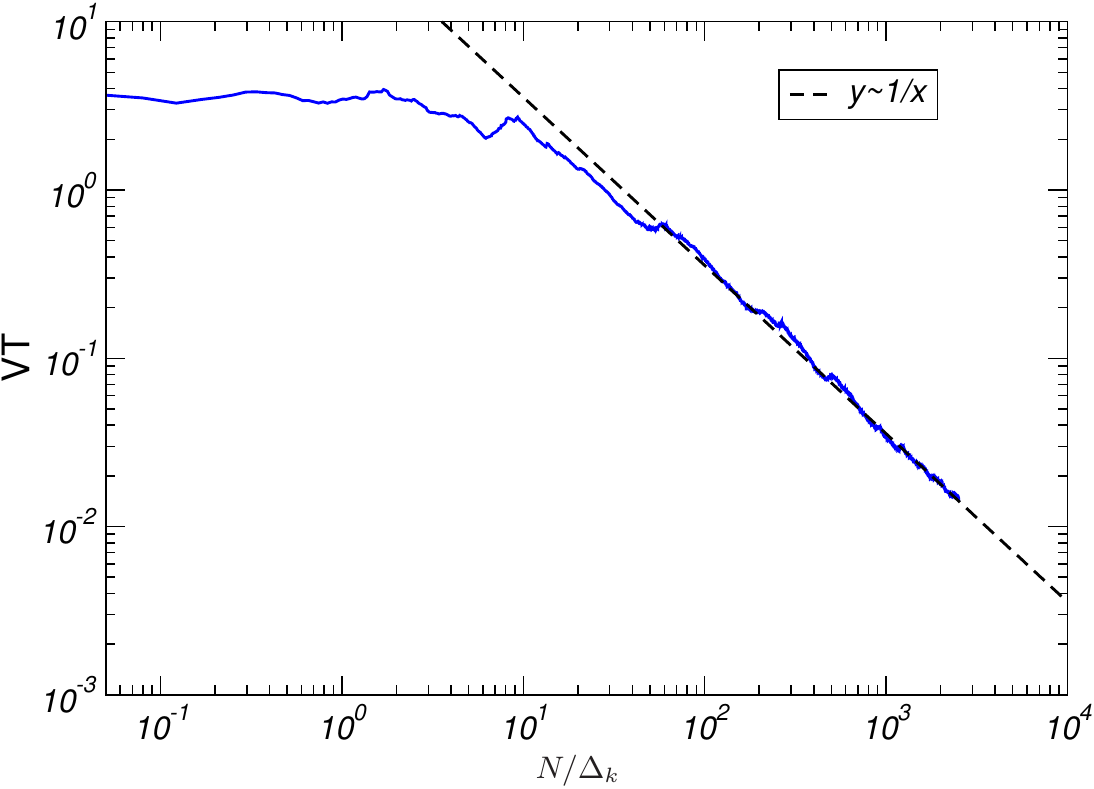}
  \caption{The variance of mean first-passage time of $160$ simulated trajectories on the sample with $\alpha=1.2$.
  }
  \label{fig:vt_quenched}
\end{figure}

The trajectory-to-trajectory fluctuation can be measured by the scaled variance of $\overline{\tau}$ by
\begin{equation}
\label{eq:vt}
\text{VT}=\frac{\left<\overline{\tau}^2\right>-\left<\overline{\tau}\right>^2}{\left<\overline{\tau}\right>^2},
\end{equation}
One can easily see it is the generalization of EB parameter for the first-passage approach, the CLT analysis employed in the above sections can be also applied here since the correlation in $\{\tau_k\}$ can be handled by the coarse-graining in space.
One may note the first-passage segment spans a region of radius $r$. The FPT $\tau$ depends only on the local diffusivity in the region. When the spanned regions of two first-passage segments denoted by $k$ and $k'$ do not share any site of the same extreme basin, saying, $\vert {\bf x}(t_{k'})-{\bf x}(t_{k})\vert>2(r+r_c)$, $\tau_k$ and $\tau_{k'}$ are uncorrelated. Noting also $\vert {\bf x}(t_{k'})-{\bf x}(t_k)\vert^2\simeq (k'-k)r^2$, one can see the correlation vanishes for $\Delta_k=k'-k>4(1+r_c/r)^2$. The CLT analysis can be then applied to the distribution of $\overline{\tau}$, which is quite similar to that for $\overline{\delta^2}$ in Sec.\ref{sec:annealed}. We show the results directly here. For $N<\Delta_k$, all the summands are correlated, since the particle scans no more than one or two extreme basin. The summation would not depress the trajectory-to-trajectory fluctuation, which reflects the fluctuation of local diffusivities on different initial sites. VT is hence kept at high level. For $N\gg\Delta_k$, the CLT suggests it vanishes as $VT\sim 1/N$. The predicted behavior is confirmed by the simulation data, as shown in Fig. \ref{fig:vt_quenched}.

\section{Discussion}
\label{sec:disc}

Two origins of the trajectory-to-trajectory variance are analysed in this study: the intrinsic stochastic feature of the random walk and the heterogeneity of the disordered environments, both the annealed and quenched cases. In the ideal case, the ergodicity would eventually recover when the self-averaging over both the origins is achieved in each trajectory. It is, however, the rare case in the experiments with limited observation time on living cells. As shown in the study, the fluctuation introduced by the disordered environments persists much longer than that by the intrinsic random feature of the walk. One may expect for the long observation, the trajectory-to-trajectory variance are mainly contributed by the heterogeneity of the media. In this sense, the variance encodes the structure information of the environments. One may utilize the information and visualize the structures by the diffusion map (see \cite{li15} for example) and other ways.

This theoretical study may provide guidance on the data analysis for the particle tracking experiments on living cells and the colloidal systems.

{\it Living cells.} Due to the heterogeneity of the cellular environments, the behaviors of diffusion in different parts of the cells varies significantly\cite{tabaka14}.
The cytoplasm of eukaryotic cells is rather dynamical\cite{guo14}. The nano-particles tracked in such systems is expected following the dynamics with fluctuating diffusivity, which has been investigated in Sec.\ref{sec:annealed}. Larger tracers are more likely to be entangled in the cellular structures, which are usually quasi-static over limited observation time. The quenched effect may arise in this case. The structures on the crowded cell membrane also relax quite slow, where the unique quenched effects have been reported\cite{he16,jeon16}.

{\it Colloidal systems.} In the colloidal systems, the tracer can be easily tracked and the environment structure also can be manipulated and imaged (see e.g. \cite{ning19}). They are hence good proving grounds for the diffusion theories. In the dense colloidal liquids, the tracer is obstructed by the colloidal particles. Since the liquid structure changes over time, the annealed disordered model may be employed in this case. ,As the counterpart, the quenched effects is expected in the static disordered colloidal matrix.

\section{Summary}
\label{sec:sum}

In this work, we study ergodicity recovery of random walk in various disordered media, which concerns how the mean of the random observable converges along the elongating trajectory to its expected value. The trajectory-wise TAMSD is chosen as the observable following the convention. The ergodicity recovery in homogeneous media is revisited in the fashion of the experiment trajectory analysis with the constraints of finite time-space resolution. It offers the first taste on how the CLT would lead to self-averaging in a series of uncorrelated random variables. In the more complicated case with the annealed dynamic heterogeneity, we show that the ergodicity recovers only when the observation time is much longer than the relaxation time of the temporal correlated diffusivity. In such case, the coarse-graining in time can cancel the correlation in the summands of the TAMSD. The CLT can then be applied, which leads to the $EB\sim 1/T$ behavior.

It has been a puzzle whether and how the ergodicity recovers in the quenched disordered media, where the whole particle population are usually split into the localized state and the mobile one. In the localized state, the particle is frozen in the area with small diffusivity, which can hardly escape the area since it walks slow. Our extensive simulation shows that the localized particles delays the ergodicity recovery for very long time, which provides insights to the slow decay of EB parameter observed in the particle tracking experiments. It also explains the abnormal TAMSD behavior previously observed in the molecular dynamics simulation (See Fig. 8 in \cite{jeon16}).

The first-passage approach is further introduced for the analysis of the trajectories in the quenched disordered media, of which the trajectory is decomposed into segments of the fixed head-to-tail distance. The ergodicity recovery analysis is generalized by choosing the FPT of the segment as the observable. Since the diffusivity is locally correlated, the CLT can be applied to the mean FPT when the space scale of the trajectory is much larger than the correlation length. The variance of the mean FPT is then depressed by $VT\sim 1/L^2$, where $L$ is the head-to-tail distance of the whole trajectory. This approach may be employed in the future analysis on the trajectories from the particle tracking experiments, especially in the case that the disordered environments is static over the experiment time scale and the particle dynamics is correlated in space but not in time.

\begin{acknowledgements}
This work is supported by National Natural Science Foundation of China (Grant No. 11705064, 11675060,  91730301). 
\end{acknowledgements}

\appendix

\section{The non-Gaussian displacement distribution in the annealed disordered case}
\label{app1}

In this appendix, we investigate the non-Gaussian diffusion in the annealed disordered media, of which case the ergodicity recovery has been studied in Sec.\ref{sec:annealed}. The study on non-Gaussian diffusion concerns the distribution of the displacement of the segments. One can start from Eq. (\ref{eq:xi}) and Eq.(\ref{eq:dx}). The head-to-tail displacement of the segment is contributed by the increments via
\begin{equation}
\label{eq:disp2}
\delta^{(x,y)}(\Delta_t)=\sum_{i=1}^{N_q}\xi^{(x,y)}_{i},
\end{equation}
where $x$ or $y$ denotes the component in the corresponding direction, and $N_q=\Delta_t/t_{\text{bin}}$.
In the short time limit $\Delta_t\ll t_{\text{D}}$, the latent variable $w$ is roughly constant for all the increments in the segment. The uncorrelated increments $\xi=(\xi^{(x)},\xi^{(y)})$ are identically distributed. The CLT then suggests
\begin{equation}
\delta^{(x,y)}(\Delta_t)\xrightarrow{d}\mathcal{N}\left(N_q\left<\xi^{(x,y)}\right>,N_q\left<\left[\xi^{(x,y)}\right]^2\right>\right),
\end{equation}
where $\left<\xi^{(x,y)}\right>=0$ and the variance $\left<\left[\xi^{(x,y)}\right]^2\right>=2wa^2t_{\text{bin}}$.
The probability density function can be explicitly written in the Gaussian form
\begin{equation}
G(\delta^{(x,y)},\Delta_t\vert w)=\frac{1}{\sqrt{4\pi wa^2 \Delta_t}}\exp\left(-\frac{\left[\delta^{(x,y)}\right]^2}{4wa^2\Delta_t}\right).
\end{equation}
Noting that the diffusivity $w$ of different segments follows the distribution given by Eq. (\ref{eq:pw}),
the displacement distribution $P(\delta^{(x,y)},\Delta_t)$ can be estimated as the marginal distribution by
\begin{equation}
P(\delta^{(x,y)},\Delta_t)=\int_0^\infty dw\;G(\delta^{(x,y)},\Delta_t\vert w)P(w).
\end{equation}
The above integral has been estimated by several approaches in previous studies\cite{sposini18,luo19}, which give the tail behavior
\begin{equation}
\label{eq:stretched1}
P(\tilde{x})\sim\frac{1}{\sqrt{2\pi}}\left(\frac{\tilde{x}^2}{2}\right)^{\frac{\alpha-1}{2(\alpha+1)}}\exp\left[-\frac{\alpha+1}{\alpha}\alpha^{\frac{1}{\alpha+1}}\left(\frac{\tilde{x}^2}{2}\right)^{\frac{\alpha}{\alpha+1}}\right],
\end{equation}
where $\tilde{x}=\delta^{(x,y)}/\sqrt{2wa^2\Delta_t}$. The stretched exponential tail is modulated by the roughness parameter $\alpha$.

\begin{figure}
\centering
  \includegraphics[width=.9\linewidth]{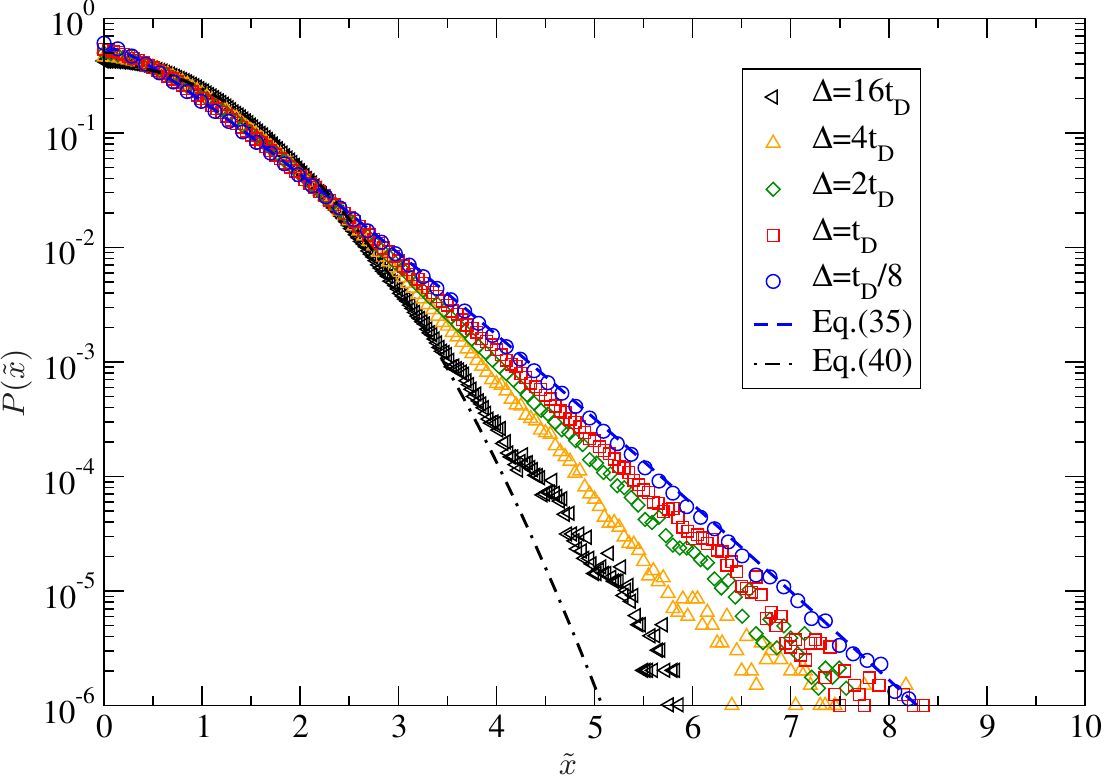}
  \caption{The rescaled displacement distribution of the annealed model with $\alpha=1.2$ and $t_{\text{D}}=200$. The lines show the limit distributions. }
  \label{fig:pdx_annealed}
\end{figure}

In the long time limit $\Delta_t\gg  t_{\text{D}}$, the diffusivity fluctuates in each segment. The summation in Eq.(\ref{eq:disp2}) can be regrouped by the time interval $ t_{\text{D}}$ as
\begin{equation}
\label{eq:disp4}
\delta^{(x,y)}(\Delta_t)=\sum_{j=1}^{N_D}\delta^{(x,y)}_j( t_{\text{D}}),
\end{equation}
where
\begin{equation}
\label{eq:disp3}
\delta^{(x,y)}_j( t_{\text{D}})=\sum_{i=1}^{N_p}\xi^{(x,y)}_{(j-1)N_p+i},
\end{equation}
and $N_D=\Delta_t/ t_{\text{D}}$ and $N_p= t_{\text{D}}/t_{\text{bin}}$. In each subgroup, the diffusivity $w(t)$ can be assumed constant. The CLT again works as
\begin{equation}
\delta^{(x,y)}_j( t_{\text{D}})\xrightarrow{d}\mathcal{N}\left(0,\sigma_j^2\right),
\end{equation}
where the variance $\sigma_j=2w_ja^2 t_{\text{D}}$ depends on the diffusivity $w_j$ during $(t,t+ t_{\text{D}})$. Their summation,  $\delta^{(x,y)}(\Delta_t)$, is then also Gaussian distributed with the zero mean and the variance
\begin{equation}
\sigma^2=\sum_{j=1}^{N_D}\sigma_j^2=2a^2 t_{\text{D}}\sum_{j=1}^{N_D} w_j.
\end{equation}
In general, $\sigma^2$ is a variable depending on the random diffusivities $w_j$. In the long time limit with $N_D\gg1$, the summation can be approximated as $\sum_{j=1}^{N_D} w_j\simeq N_D\left<w\right>$. All the $\delta^{(x,y)}(\Delta_t)$ have almost the same variance. The distribution of $\delta^{(x,y)}(\Delta_t)$ then converges to a pure Gaussian distribution,
\begin{equation}
\label{eq:gaussian}
P(\tilde{x})=\frac{1}{\sqrt{2\pi}}\exp\left(-\frac{\tilde{x}^2}{2}\right).
\end{equation}
Eq. (\ref{eq:stretched1}) and Eq. (\ref{eq:gaussian}) are plotted in Fig. \ref{fig:pdx_annealed}. One can see the two limit distributions well contain the simulation results.

\section{The simulation details}
\label{app2}
This appendix provides the details of the simulation procedure.

The trap dynamics is simulated by the Gillespie algorithm\cite{gillespie77}, which interprets the random walk as a series of stochastic events with random time intervals. In the homogeneous case, it is composed of the jumping events on the lattice.
The waiting time between two jumps is generated from the exponential distribution
\begin{equation}
\label{eq:wt}
P(\tau)=W_{\text{total}}\exp(-W_{\text{total}}\tau),
\end{equation}
where $W_{\text{total}}=4w$ is the total transition rate to the four nearest-neighboring sites.
The direction of each jump is randomly chosen with even probability. $w$ is set to $1$ in the simulation. The trajectory is discretized as described in Sec. \ref{sec:sa}, where $t_{\text{bin}}=10$.

In the annealed disordered case with fluctuating diffusivity,  the process is defined as a series of position and diffusivity, $\{{\bf x}_i,w_i\}$. The stochastic events are then constituted by the jumps on the lattice and the diffusivity resampling from the distribution given by Eq.(\ref{eq:pw}). According to the Gillespie algorithm, the waiting time between two events also follows the distribution given by Eq.(\ref{eq:wt}), while $W_{\text{total}}=4w_i+w_D$ is the total transition rate including the jumping rate $w_i$ and the diffusivity resampling rate $w_D$. Each event can be either a jump to one of the neighbour sites with the probability $w_i/W_{\text{total}}$ or the diffusivity resampling with the probability $w_D/W_{\text{total}}$. The typical relaxation time of the diffusivity is then given as $t_{D}=1/w_{D}$.

In the quenched disordered case with spatial correlated diffusivity, the static disordered landscape is first generated by a two-step procedure shown in Sec.\ref{sec:quenched}. The size of the disordered landscape $\{v_i\}$ are chosen as $L_x=L_y=1024$, while the radius of the extreme basin is set as $r_c=16$. The periodic boundary condition is applied for long time simulation. The initial site of each trajectories is randomly chosen following the Boltzmann distribution. The Gillespie algorithm for the random walk is similar to the case of the homogeneous case but with the site-dependent diffusivity $w_i=w_0\exp(v_i/\alpha)$.

\FloatBarrier



\begin{thebibliography}{99}


\bibitem{li15} H.~Li, S.-X. Dou, Y.-R. Liu, W.~Li, P.~Xie, W.-C. Wang and P.-Y. Wang, J. Am. Chem. Soc. \textbf{137}, 436 (2015).

\bibitem{he16} W.~He, H.~Song, Y.~Su, L.~Geng, B.~J.~Ackerson, H.~B.~Peng and P.~Tong, Nat. Commun. \textbf{7}, 11701 (2016).

\bibitem{munder16} M.~C.~Munder et al., eLife \textbf{5}, e09347 (2016).

\bibitem{li18} B.~Li, S.~X.~Dou,, J.~W.~Yuan, Y.~R.~Liu, W.~Li, F.~Ye, P.-Y.~Wang and H.~Li, Proc.~Natl.~Acad.~Sci.~U.S.A \textbf{ 115}, 12118 (2018).

\bibitem{ning19} L.~Ning, P.~Liu, Y.~ Zong, R.~Liu, M.~Yang and K.~Chen, Phys. Rev. Lett. \textbf{122} 178002 (2019). 

\bibitem{sentjabrskaja16} T.~Sentjabrskaja {\it et al.}, Nature Comm. \textbf{7} 11133 (2016).

\bibitem{kou17} B.~Kou, et al., Nature \textbf{551}, 360 (2017).

\bibitem{manzo15} C.~Manzo, J.~A.~Torreno-Pina, P.~Massignan, G.~J.~Lapeyre, Jr. , M.~Lewenstein and M.~F.~Garcia Parajo, Phys. Rev. X \textbf{5}, 011021 (2015).

\bibitem{luo18} L.~Luo and M.~Yi, Phys.~Rev.~E \textbf{ 97}, 042122 (2018).

\bibitem{barkai12} E.~Barkai, Y.~Garini and R.~Metzler, Phys. Today \textbf{65} 29 (2012).

\bibitem{metzler14} R.~Metzler, J.-H.~Jeon, A.~G.~Cherstvy and E.~Barkari, Phys.~Chem.~Chem.~Phys.~{\bf 16} 24128 (2014). 

\bibitem{montroll75} H.~Scher and E.~W.~Montroll, Phys. Rev. B \textbf{12} 2455 (1975). 

\bibitem{luo14} L.~Luo and L.-H.~Tang, Chin. Phys. B \textbf{23}, 070514 (2014). 

\bibitem{bel05} G.~Bel and E. Barkai, Phys. Rev. Lett. \textbf{94}, 240602 (2005).

\bibitem{he08} Y.~He, S.~Burov, R.~Metzler and E.~Barkai, Phys.~Rev.~Lett.~{\bf 101}, 058101 (2008). 

\bibitem{wang09} S.~C.~Bae, B.~Wang, J.~Guan and S.~Granick, Proc.~Natl.~Acad.~Sci. \textbf{106}, 15160 (2009).

\bibitem{liu17} J.~Liu, B.~H.~Li and X. S. Chen, Chin.~Phys.~Lett., \textbf{34} 050201 (2017).

\bibitem{jeon16} J.-H.~Jeon, M.~Javanainen, H.~Martinez-Seara, R.~Metzler and I.~Vattulainen, Phys. Rev. X \textbf{6}, 021006 (2016).

\bibitem{cherstvy13} A.~Cherstvy and R.~Metzler, Phys.~Chem.~Chem.~Phys.~{\bf 15}, 20220 (2013).

\bibitem{luo19} L.~Luo and M.~Yi, Phys. Rev. E \textbf{100} 042136 (2019).

\bibitem{guo18} W.~Guo, Y.~Li, W.-H.~Song and L.-C.~Du, J.~Stat.~Mech.~ 033303 (2018). 

\bibitem{machta81} J.~Machta, Phys.~Rev.~B \textbf{24}, 5260 (1981).

\bibitem{haus87} J.~W.~Haus and K.~W.~Kehr, Phys.~Rep. \textbf{150}, 263 (1987).

\bibitem{jeon14} J.-H.~Jeon, A.~V.~Chechkin and R.~Metzler, Phys.~Chem.~Chem.~Phys.~{\bf 16} 15811 (2014).

\bibitem{bouchaud90} J.~P.~Bouchaud and A.~Georges, Phys.~Rep. \textbf{ 195}, 127 (1990).

\bibitem{luo15} L.~Luo and L.-H.~Tang, Phys.~Rev.~E \textbf{ 92}, 042137 (2015).

\bibitem{slater14} M.~V.~Chubynsky and G.~W.~Slater, Phys. Rev. Lett. \textbf{113}, 098302 (2014).

\bibitem{tabaka14} M.~Tabaka, T.~Kalwarczyk, J.~Szymanski, S.~Hou and R.~Holyst, Front. Phys. \textbf{2}, 54 (2014).

\bibitem{guo14} M.~Guo, A.~J.~Ehrlicher, M.~H.~Jensen, M.~Renz, J.~R.~Moore, R.~D.~Goldman, J.~Lippincott-Schwartz, F.~C.~Mackintosh and D.~A.~Weitz, Cell \textbf{158}, 822 (2014). 

\bibitem{sposini18} V.~Sposini, A.~V.~Chechkin, F.~Seno, G.~Pagnini, and R.~Metzler, New~J.~Phys. \textbf{20}, 043044 (2018).

\bibitem{gillespie77} D.~T.~Gillespie, J.~Phys.~Chem. \textbf{ 81}, 2340 (1977).



\end{thebibliography}
\end{document}